# Observation of Weak Temperature Dependence of Spin Diffusion Length in Highly-doped Si by Using a Non-local 3-terminal Method


M. Kameno*, E. Shikoh, T. Oikawa[1], T. Sasaki[1], T. Suzuki[2], Y. Suzuki and M. Shiraishi

Graduate School of Engineering Science, Osaka Univ. Osaka, Japan

[1]SQ Research Center, TDK Corporation, Nagano, Japan

[2]AIT, Akita Research Institute of Advanced Technology, Akita, Japan


## Abstract


We conduct an experimental investigation of temperature dependence of spin diffusion length in highly-doped n-type silicon by using a non-local 3-terminal method. Whereas an effect of spin drift is not negligible in non-metallic systems, it is not fully conclusive how the spin drift affects spin transport properties in highly-doped Si in a non-local 3-terminal method. Here, we report on temperature dependence of spin diffusion length in the Si, and it is clarified that the spin transport is less affected by an external electric field.




Si spintronics has attracted much attention in the last decade, since Si possesses a comparatively small spin orbit interaction due to its lattice inversion symmetry and its low atomic number. Hence, it is widely recognized that Si spintronics opens a new research field related with the so-called beyond-CMOS devices. Furthermore, Si has no toxicity and is an environmentally material. Up to now, many studies in Si spintronics have been intensively conducted by using a hot electron transistor structure [1], non-local 4-terminal [2,3,4] and 3-terminal [5] methods and a spin-pumping method [6,7]. Although spin transport up to room temperature (RT) [3] and accumulation up to 500 K [8] were achieved to date, the detailed spin transport properties in a Si channel have not been fully understood yet because the number of studies on simultaneous realization of spin injection and transport up to RT are still limited. In this study, we introduce a lateral Si spin valve, where the non-local spin transport is realized up to RT, and report on the temperature dependence of the spin transport properties under a dc-biased condition from a viewpoint of spin drift.

A lateral Si spin valve in this study consists of a P-doped ($\sim 5 \times 10^{19}$ cm$^{-3}$) n-type Si channel with a width and thickness of 21 μm and 80 nm, respectively, on a SOI



substrate, equipped with two ferromagnetic (FM1 and FM2) and two nonmagnetic electrodes (NM1 and NM2), as shown in Fig.1. The detail of the sample preparation method is described in ref. [3]. A non-local 4-terminal measurement [3] allowed us to observe the spin signal in this device up to RT (not shown here). The material of the FM electrodes is Fe onto an MgO tunneling barrier layer. The width of FM2 was 2.0 μm, and the NM electrodes were made of Al. In this study, we do not use FM1 electrodes in non-local 3-terminal measurements because the width of FM1 electrode (0.45 μm) is shorter than the intrinsic spin diffusion length in n-Si at 8 K (approximately 1.8 μm estimated by observing a 4-terminal Hanle effect, see ref. [4]), because precise estimation of spin coherence by using the non-local 3-terminal method is difficult if the width of an FM electrode is shorter than spin diffusion length as already discussed in ref. [5]. The 3-terminal Hanle-effect measurement [5] provides a simple means of demonstrating the spin accumulation in Si just below FM2 by applying an external magnetic field parallel to the z-axis. An electric current was applied to the direction from FM2 to NM1, which is defined as the direction of upstream spin flow [9].

Figure 2 (a) shows the 3-terminal Hanle voltages as a function of the magnetic field



from -200 to 200 Oe at 8 K, 50 K, 100 K, and 200 K, where an injection current was set

to be -4.0 mA, which corresponds to the bias voltage of 1.84 V. A parabolic background

depended on the magnetic field was subtracted from the raw data. It is obvious that the

Hanle-type spin precession of the accumulated spins was detected under this condition.

It is noteworthy that the Hanle signals were not observed under the positive bias

currents, which will be discussed elsewhere [10]. The spin signal monotonically

decreased with temperature as shown in Fig. 2(a). Here, we estimate spin lifetime, $\tau$,

using the following fitting function,

$$
\begin{aligned}
\frac{V(B)}{I} = \frac{V_0}{I} &\exp\left(-\frac{d}{\lambda}\right)(1+\omega^2\tau^2)^{-\frac{1}{4}} \\
&\times \exp\left[\frac{-d}{\lambda}\left(\sqrt{\frac{\sqrt{1+\omega^2\tau^2}+1}{2}}-1\right)\right] \\
&\times \cos\left[\sqrt{\frac{\left(\tan^{-1}(\omega\tau)\right)^2}{2}}+\frac{d}{\lambda}\sqrt{\frac{\sqrt{1+\omega^2\tau^2}-1}{2}}\right],
\end{aligned}
\tag{1}
$$

where $d$ is the center-to-center length of the two ferromagnetic electrodes, $\lambda$ is the spin

diffusion length, $\omega = g\mu_B B/\hbar$ is the Lamor frequency, $g$ is the electron g-factor ($g$ =2),

$\mu_B$ is the Bohr magneton [11]. It is noted that the fitting function is an analytical

solution of a conventional Hanle-type spin precession [12], where the distance between

the ferromagnetic electrodes was assumed to be 0 in the fitting function in the



4-terminal method. Equation (1) can fit the experimental data well (see Fig. 2(b)), and

the estimated $\tau$ values are almost constant at low temperature and becomes smaller

down to 2 ns at 200 K as shown in Fig. 3(a). The relationship between the spin diffusion

length, $\lambda$, and the spin lifetime is described as $\lambda = \sqrt{D\tau}$. Here, $D$ is the diffusion

constant and was already estimated to be 4.3 cm²/s by using the non-local 4-terminal

method at 8 K [3]. The spin diffusion length in a semiconductor is known to be

modulated by an electric field and temperature, which has been clarified by solving the

spin drift-diffusion equation [13],

$$\nabla^2 \left(n_\uparrow - n_\downarrow\right) - \frac{eE}{k_B T} \cdot \nabla\left(n_\uparrow - n_\downarrow\right) - \frac{1}{\lambda^2}\left(n_\uparrow - n_\downarrow\right) = 0, \qquad (2)$$

where $n_\uparrow - n_\downarrow$ is the difference between up-spin and down-spin electron densities, $k_B$ is

the Boltzmann constant, $T$ is temperature, $e$ is the electric charge, $E$ is an electric field,

and the general form of the upstream spin diffusion length, $\lambda_u$, was found to be,

$$\lambda_u{}^{-1} = \frac{\mid eE \mid}{2k_B T} + \sqrt{\left(\frac{eE}{2k_B T}\right)^2 + \frac{1}{\lambda^2}}. \qquad (3)$$

The experimentally obtained data are fitted to eq. (3) if the spin drift in this

experimental set-up is not negligible as in the case of non-degenerate semiconductors.

However, Fig.3 (b) obviously shows that the theoretically predicted values (a green solid



line) obtained by using eq. (3) does not fit the experimental result (red closed circles), namely, the spin transport was less affected by the temperature. This finding is attributed to the fact that the spin channel was highly-doped and spin drift contributed to the spin transport very weakly. In fact, the spin diffusion length is constant with temperature in the model for a degenerate semiconductor and our study is the case, whereas the spin diffusion length slightly decreased at high temperature, which is likely due to the decrease of spin polarization [3]. The spin diffusion length in the model for a degenerate semiconductor is rewritten by using Einstein's relation, $D/\mu = 2E_F/3e$, as

$$\lambda_u^{-1} = \frac{|3eE|}{4E_F} + \sqrt{\left(\frac{|3eE|}{4E_F}\right)^2 + \frac{1}{\lambda^2}}, \qquad (4)$$

where $D$ is the diffusion constant and $\mu$ is the mobility. $E_F$ is the Fermi energy of the silicon and its value at 8 K was estimated to be 50 meV [14]. As shown in Fig. 3(b), the spin diffusion length (a blue solid line) in a degenerate semiconductor model was calculated by using eq. (4) and assuming that the Fermi energy is temperature-independent, which fits the experimental data (see the blue solid line in Fig. 3(b)). Although several groups have independently investigated the temperature



dependence of the spin accumulation voltages in the 3-terminal method, it is notable that this is the first observation of the dependence in the lateral Si spin valve where the spin transport was realized at RT. The dependence observed is similar to those in the highly-doped Si [8,13] and is not, in contrast, similar with that from the sample without Cs-doping in the Si channel [5].

In conclusion, we investigated the temperature dependence of the spin accumulation signals in the lateral Si spin valves, where spin transport was realized up to RT. It was clarified that the spin diffusion length in the highly-doped Si was less dependent on temperature by using the non-local 3-terminal method as was expected because the spin drift does not strongly contribute to the spin transport.




Reference

[1]  I. Appelbaum, Biqin Huang, and D. J. Monsma: Nature **477** (2007) 17.

[2]  T. Sasaki, T. Oikawa, T. Suzuki, M. Shiraishi, Y. Suzuki, and K. Tagami: Appl. Phys. Express **2** (2009) 053003.

[3]  T. Suzuki, T. Sasaki, T. Oikawa, M. Shiraishi, Y. Suzuki, and K. Noguchi: Appl. Phys. Express **4** (2011) 023003.

[4]  M. Shiraishi, Y. Honda, E. Shikoh, Y. Suzuki, and T. Shinjo: Phys. Rev. B **83**, (2011) 241204(R).

[5]  S. P. Dash, S. Sharma, R. S. Patel, M. P. de Jong and Ron Jansen: Nature **462** (2009) 26.

[6]  E. Shikoh, K. Ando, E. Saitoh, T. Shinjo, M. Shiraishi: arXiv:1107.0376.

[7]  K. Ando, E. Saitoh: arXiv:1107.2585.

[8]  C.H. Li, O.M.J. van 't Erve, and B.T. Jonker: Nature Communications **2** (2011) 245.

[9]  Z. G. Yu and M. E. Flatté: Phys. Rev. B **66** (2002) 201202(R).

[10] M. Kameno et al., in preparation.

[11] T. Sasaki, T. Oikawa, T. Suzuki, M. Shiraishi, Y. Suzuki, and K. Noguchi: IEEE





Trans. Magn. **46** (2010) 1436.

[12] F. J. Jedema, A. T. Filip and B. J. van Wees: Nature **410** (2001) 345.

[13] K. Jeon, B. Min, I. Shin, C. Park, H. Lee, Y. Jo, and S. Shin: Appl. Phys. Lett. **98**, (2011) 262102.

[14] The conductivity and the diffusion constant of the Si channel at 8 K were measured to 1.03×10$^5$ Ω$^{-1}$ m$^{-1}$ and 4.3 cm$^2$/s, respectively, and the mobility was calculated to be 129 cm$^2$/Vs by using the value of the conductivity. Then, the Fermi energy was estimated to be 50 meV by using the Einstein's relation, $D/\mu = 2E_F/3e$.




Figure captions

Fig.1. A schematic of the lateral Si spin device (not to scale). The highly doped ($5 \times 10^{19}$ cm$^{-3}$) Si channel is equipped with four electrodes (two NM, two FM). An electric current is injected into one circuit (NM1-Si-FM2), and an output spin accumulation voltage is detected by the other circuit (FM2-Si-NM2). FM1 electrode is not used in this study. The external magnetic field was applied to the $z$ axis in the non-local 3-terminal Hanle-type spin precession measurements at 8 K, 50 K, 100 K and 200 K, respectively.

Fig.2. (a) Hanle-type spin precession properties observed at 8 K, 50 K, 100 K and 200 K. The injection current was set to be -4.0 mA. (b) Comparison between the experimental data at 50 K and the theoretical fitting curve for the Hanle-type spin precession. The red open circles are experimental data, and the red solid line is the fitting curve. The $\tau$ at 50 K was estimated to be 4.1 ns.

Fig.3. (a) Temperature dependence of the spin lifetime. The red closed circles are the results obtained by using eq. (1). (b) Temperature dependence of the spin diffusion length. The red closed circles are the estimated values of the spin diffusion length. The



green and blue solid lines show the theoretically predicted trend obtained by using eq.

(3), which are based on non-degenerate and degenerate semiconductor models, respectively.



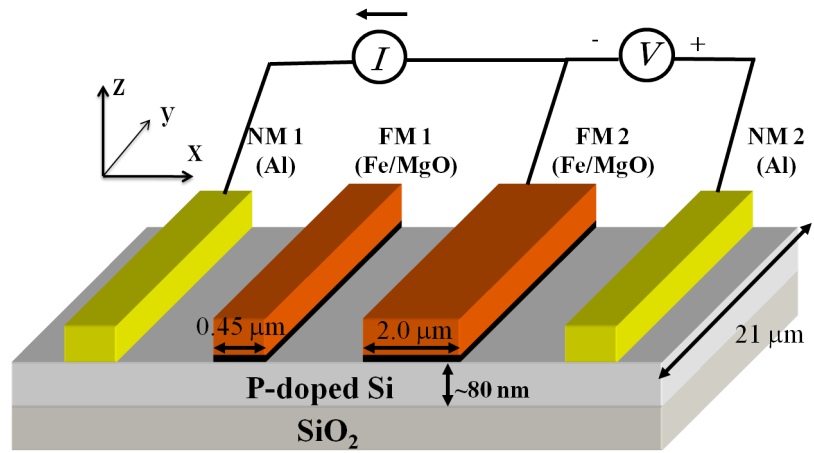

Kameno, et. al., Fig. 1.



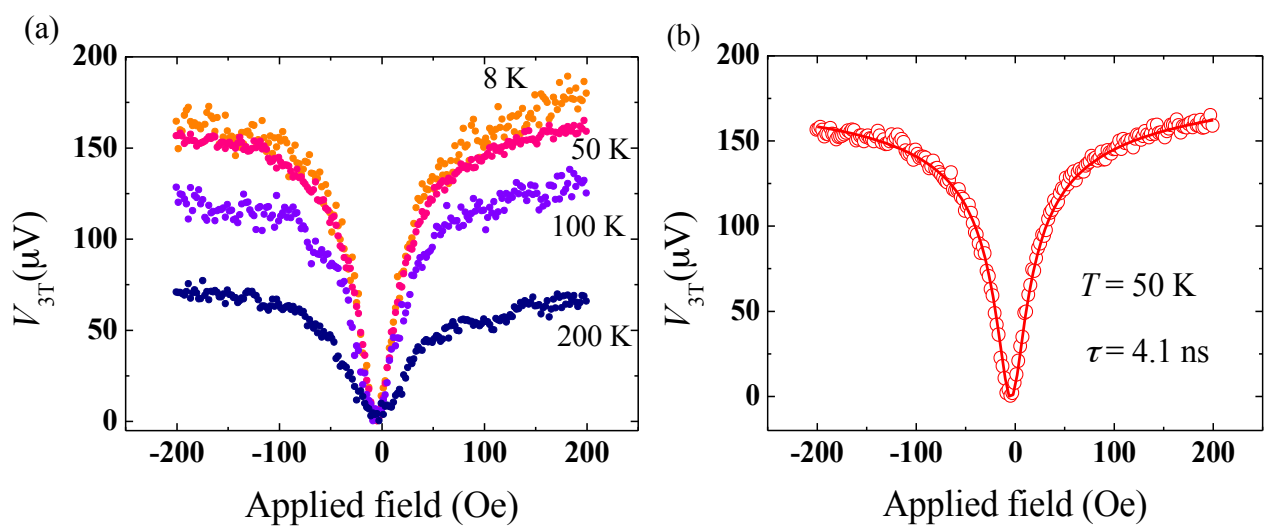



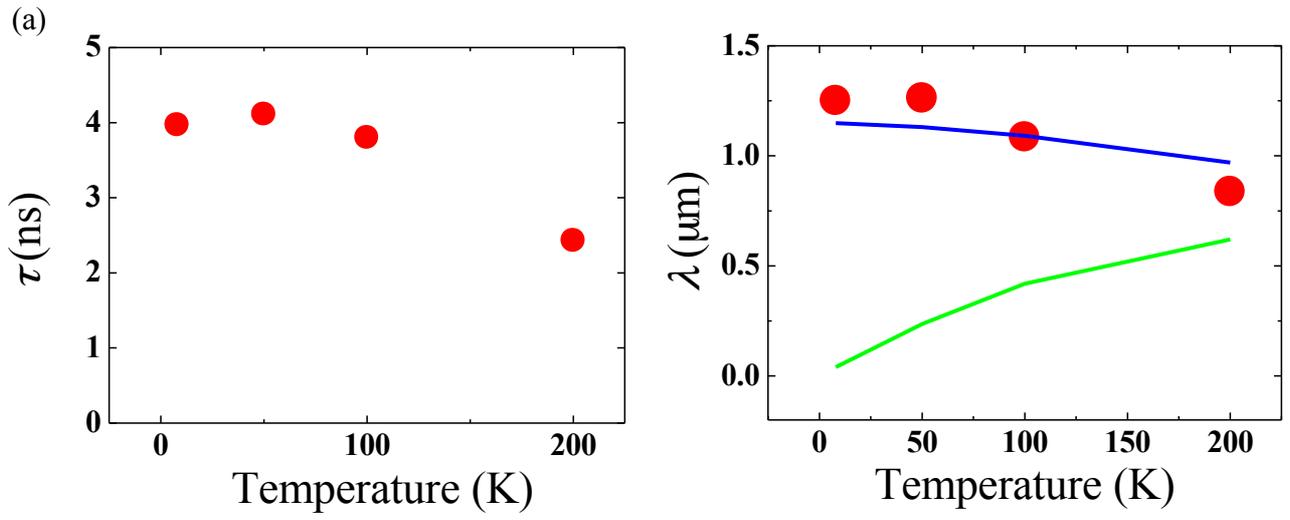

(a)

Kameno, et. al., Fig.3.